\documentclass[twocolumn,
superscriptaddress,prl,floatfix,footinbib]{revtex4-2}
\usepackage[]{graphicx}
\usepackage[usenames,dvipsnames]{color}
\usepackage{soul}
\usepackage{bm}

\usepackage{times}
\usepackage{mathrsfs}
\usepackage{graphicx}
\usepackage{dcolumn}
\usepackage{bm}
\usepackage{color}

\usepackage[colorlinks,bookmarks=false,citecolor=blue,linkcolor=red,urlcolor=blue]{hyperref}
\usepackage{multirow}

\usepackage{siunitx}
\mathchardef\mhyphen="2D
\usepackage[libertine,bigdelims,vvarbb,scaled=1.05]{newtxmath} 
 
\usepackage{babel}
\usepackage[babel=true]{microtype}

\begin{document}

\title{Observation of room temperature excitons in an atomically thin topological insulator}

\author{Marcin Syperek}\thanks{These authors contributed equally: Marcin Syperek, Raul St\"uhler and Armando Consiglio.}
\affiliation{Department of Experimental Physics, Faculty of Fundamental Problems of Technology, Wroc\l aw University of Science and Technology, Wybrze\.ze Wyspia\'nskiego 27, 50-370 Wroc\l aw, Poland}

\author{Raul St\"uhler}\thanks{These authors contributed equally: Marcin Syperek, Raul St\"uhler and Armando Consiglio.}
\affiliation{Physikalisches Institut and W\"urzburg-Dresden Cluster of Excellence ct.qmat, Universit\"at W\"urzburg, 97074 W\"urzburg, Germany}

\author{Armando Consiglio}\thanks{These authors contributed equally: Marcin Syperek, Raul St\"uhler and Armando Consiglio.}
\affiliation{Institut f\"{u}r Theoretische Physik und Astrophysik and W\"{u}rzburg-Dresden Cluster of Excellence ct.qmat, Universit\"{a}t W\"{u}rzburg, 97074 W\"{u}rzburg, Germany}

\author{Pawe{\l} Holewa}
\affiliation{Department of Experimental Physics, Faculty of Fundamental Problems of Technology, Wroc\l aw University of Science and Technology, Wybrze\.ze Wyspia\'nskiego 27, 50-370 Wroc\l aw, Poland}

\author{Pawe{\l} Wyborski}
\affiliation{Department of Experimental Physics, Faculty of Fundamental Problems of Technology, Wroc\l aw University of Science and Technology, Wybrze\.ze Wyspia\'nskiego 27, 50-370 Wroc\l aw, Poland}

\author{{\L}ukasz Dusanowski}
\affiliation{Department of Experimental Physics, Faculty of Fundamental Problems of Technology, Wroc\l aw University of Science and Technology, Wybrze\.ze Wyspia\'nskiego 27, 50-370 Wroc\l aw, Poland}
\affiliation{Physikalisches Institut and W\"urzburg-Dresden Cluster of Excellence ct.qmat, Universit\"at W\"urzburg, 97074 W\"urzburg, Germany}

\author{Felix Reis}
\affiliation{Physikalisches Institut and W\"urzburg-Dresden Cluster of Excellence ct.qmat, Universit\"at W\"urzburg, 97074 W\"urzburg, Germany}

\author{Sven H\"ofling}
\affiliation{Physikalisches Institut and W\"urzburg-Dresden Cluster of Excellence ct.qmat, Universit\"at W\"urzburg, 97074 W\"urzburg, Germany}

\author{Ronny Thomale}
\affiliation{Institut f\"{u}r Theoretische Physik und Astrophysik and W\"{u}rzburg-Dresden Cluster of Excellence ct.qmat, Universit\"{a}t W\"{u}rzburg, 97074 W\"{u}rzburg, Germany}

\author{Werner Hanke}
\affiliation{Institut f\"{u}r Theoretische Physik und Astrophysik and W\"{u}rzburg-Dresden Cluster of Excellence ct.qmat, Universit\"{a}t W\"{u}rzburg, 97074 W\"{u}rzburg, Germany}

\author{Ralph Claessen}\email{claessen@physik.uni-wuerzburg.de}
\affiliation{Physikalisches Institut and W\"urzburg-Dresden Cluster of Excellence ct.qmat, Universit\"at W\"urzburg, 97074 W\"urzburg, Germany}

\author{Domenico Di Sante}\email{domenico.disante@unibo.it}
\affiliation{Department of Physics and Astronomy, University of Bologna, 40127 Bologna, Italy}
\affiliation{Center for Computational Quantum Physics, Flatiron Institute, New York, NY 10010, USA}

\author{Christian Schneider}\email{christian.schneider@uni-oldenburg.de}
\affiliation{Institute of Physics, University of Oldenburg, 26129 Oldenburg, Germany}

\date{\today}


\begin{abstract} 
\bf{Optical spectroscopy of ultimately thin materials has significantly enhanced
our understanding of collective excitations in low-dimensional semiconductors.
This is particularly reflected by the rich physics of excitons in atomically
thin crystals which uniquely arises from the interplay of strong Coulomb
correlation, spin-orbit coupling (SOC), and lattice geometry. Here we extend the
field by reporting the observation of room temperature excitons in a material of
non-trivial global topology. We study the fundamental optical excitation
spectrum of a single layer of bismuth atoms epitaxially grown on a SiC substrate
(hereafter bismuthene or Bi/SiC) which has been established as a large-gap,
two-dimensional (2D) quantum spin Hall (QSH) insulator. Strongly developed
optical resonances are observed to emerge around the direct gap at the K and K'
points of the Brillouin zone, indicating the formation of bound excitons with
considerable oscillator strength. These experimental findings are corroborated,
concerning both the character of the excitonic resonances as well as their
energy scale, by ab-initio \emph{GW} and Bethe-Salpeter equation calculations,
confirming strong Coulomb interaction effects in these optical excitations. Our
observations provide evidence of excitons in a 2D QSH insulator at
room temperature, with excitonic and topological physics deriving from the very
same electronic structure.}
\end{abstract}

\maketitle

\section*{Introduction}


Optical spectroscopy conducted
on atomically thin sheets of transition metal dichalcogenides (TMDCs) has opened
a broad spectrum of fundamental research, and significantly enhanced our
understanding of the physics of elementary excitations in low-dimensional
semiconductors~\cite{mak2010atomically,mak2018light}.
Indeed, the dramatic increase of Coulomb correlations in two-dimensional structures has 
revealed the emergence of a vast variety of many-body states, including
non-hydrogenic Rydberg series of excitons~\cite{chernikov2014exciton}, stable
charged-~\cite{mak2013tightly, ross2013electrical}, and multi-excitonic 
complexes~\cite{you2015observation, wang2018colloquium}.

Our observation of excitons in a large-gap QSH insulator at room
temperature opens up a new exciting avenue of combining excitonic physics and
topological electronic properties. Here a first central step was recognizing
that the optical selection rules are governed by the winding number of the
low-energy gapped Dirac bandstructure, which like the Berry curvature, is a
``local" topological quantity in 
{\bf $k$}-space~\cite{Mak2018,cao2012valley, zeng2012valley,srivastava2015signatures}.
This local physics is indeed determined by the Berry curvature flux through
the small {\bf $k$}-space spanned by the relative motion of the $e-h$ pair in
the excitonic wavefunction~\cite{ZhouPRL115}. This appears, {\it e.g.}, in a
class of materials called gapped chiral fermion systems, which comprises gapped
topological surface states~\cite{Garate2011}, monolayers of
TMDCs~\cite{XiaoPRL2012} and biased bilayers with broken inversion
symmetry~\cite{McCann2006}. Recent work indeed has connected the strength and
required light polarization of an excitonic transition with the corresponding
optical matrix elements' winding number~\cite{Zhang2018,Cao2018,Xu2020}, and
thus successfully described the canonical valley-contrasting selection rules
of K and K' excitons in TMDCs.

However, despite the fact that optical properties of TMDCs thus are strongly
impacted by the above local {\bf $k$}-space phenomena ~\cite{Mak2018}, the
commonly investigated pristine monolayer TMDCs WS$_2$, WSe$_2$, MoS$_2$,
MoSe$_2$, and MoTe$_2$, in their 2H-phase, belong to the class of topologically
trivial insulators. From a fundamental perspective, the implications of the
interlink between optical selection rules and topological band invariants (a
direct link between excitonic physics and global topology, {\it i.e.}, the
Chern number as {\bf $k$}-space integral over the Berry curvature) are expected
to trigger a plethora of future experimental and theoretical investigations.
First steps in this direction are discussed in our conclusions and in the Supplementary Information.


\begin{figure*}[!t]
\centering
\includegraphics[width=\textwidth,angle=0,clip=true]{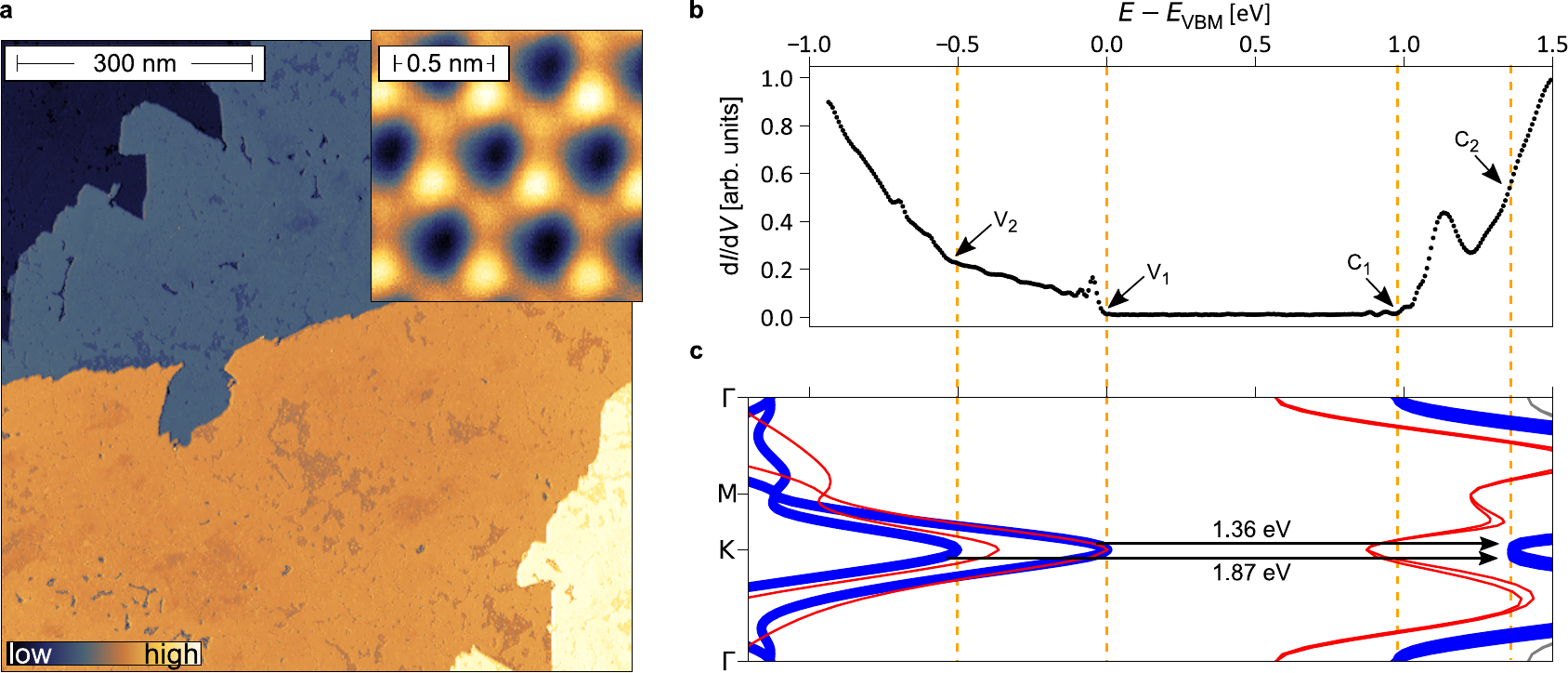}
\caption{{\bf Single-particle electronic properties of Bi/SiC.} 
{\bf a} 
Constant current STM image of bismuthene on SiC(0001).
The topographic map displays the high coverage ($\sim87\%$) of the epitaxially
grown Bi monolayer and the terrace structure of the underlying SiC substrate. ($T=\SI{4.2}{\kelvin}$ and
$V_{\text{set}} = \SI{2.6}{\volt}$, $I_{\text{set}} = \SI{50}{\pico\ampere}$.)
Inset: A high-resolution measurement reveals the bismuthene honeycomb lattice.
($T=\SI{4.2}{\kelvin}$ and $V_{\text{set}} = \SI{-0.4}{\volt}$, $I_{\text{set}}
= \SI{200}{\pico\ampere}$.)
{\bf b} 
Differential tunneling conductivity $(\mathrm{d}I/\mathrm{d}V)$ spectrum, 
locally measured in the center of the bismuthene
film ($T=\SI{4.2}{\kelvin}$ and $V_{\text{set}} = \SI{-0.4}{\volt}$, 
$I_{\text{set}} = \SI{200}{\pico\ampere}$, $V_{\text{mod}} = \SI{10}{\milli\volt}$). 
Note that the peak at $\approx 1.1$ eV is not 
an intrinsic feature of the bismuthene film, but originates from a local defect
state in the SiC substrate (see Section II of the Supplementary Information).
{\bf c} 
$GW$ band structure calculation (blue curves) for bismuthene. The direct 
optical transitions at the K-point are indicated by arrows. Also shown is the DFT band
structure obtained with the HSE hybrid functional (red
curves)~\cite{Reis2017}, illustrating the propensity of DFT to
underestimate the band gaps.
Characteristic points in the spectrum that correspond to
the valence band onsets ($V_1$, $V_2$) and the conduction band onsets ($C_1$,
$C_2$) are marked.}
\label{fig1}
\end{figure*}

To date, though, any experimental scrutiny of the impact of non-trivial global band topology on
light-matter interaction and exciton physics has so far been hampered by the
lack of suitable topological insulators with sufficiently wide band gap to allow
coupling to high-energy photons. It is only very recently that such 2D topological
insulators have been discovered and synthesized, namely 1T'-WTe$_2$
monolayers~\cite{Wu2018} and Bi/SiC, an atomically thin bismuth layer, which
is arranged in a graphene-like unbuckled honeycomb lattice~\cite{Reis2017}. In bismuthene, the
SOC induced by the heavy Bi atoms and the orbital filtering resulting from
the covalent bonding to the substrate are responsible for the emergence of a
giant topological band-gap~\cite{GangLi2018}, which is large
enough to be probed by optical spectroscopy in the near infrared. This is in
striking contrast to any other topological insulator previously known, and opens
up novel experimental opportunities to the study of photo-induced excitonics 
in 2D topological insulators.

Here, we study the electronic as well as the optical band-gap of Bi/SiC
employing photo-modulated reflectivity and scanning tunnelling spectroscopy
(STS) combined with ab-initio $GW$ calculations, accounting for the electronic Coulomb
interactions~\cite{Hedin1965,StrinatiPRL1980,Strinati1982}. Our investigation reveals the
emergence of strong optical resonances within the electronic band-gap which we
associate with the material's excitonic modes. The experimentally derived
excitonic resonance energies are notably well reproduced by many-body
calculations in the $GW$+Bethe--Salpeter equation
framework~\cite{Hanke1975,Hanke1979,HankeShamPRB1980,Rohlfing1998,Onida2002}. Our findings, which 
evidence the emergence of excitons at room temperature in a large-gap 
QSH insulator, ideally advance present activities to link 
excitonics and topological materials research
~\cite{Zhang2018,Cao2018,Xu2020,Seradjeh2009,Pikulin2014,Budich2014,
Fuhrman2015,Du2017,MacDonald2017,Chen2017,Blason2020,Varsano2020}.


\begin{figure*}
\centering
\includegraphics[width=\textwidth,angle=0,clip=true]{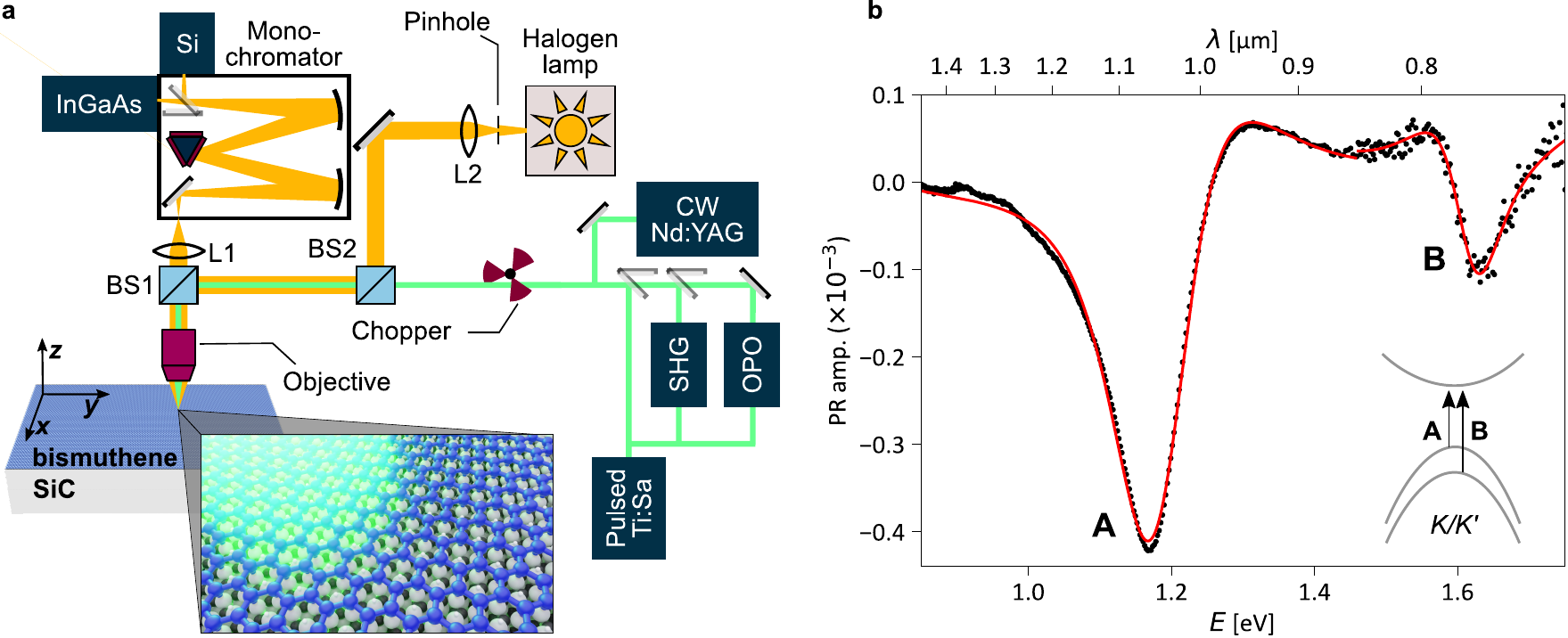}
\caption{{\bf Optical spectroscopy of Bi/SiC.} 
{\bf a} 
The experimental setup for high-spatial-resolution photo-modulated reflectivity
(PR). 
{\bf b} 
The fundamental excitation spectrum of bismuthene measured in the PR experiment
at $T=\SI{300}{\kelvin}$ (CW excitation, $E_{\mathrm{laser}}\sim\SI{2.33}{\electronvolt}$,
$P_{\mathrm{ave}}\sim\SI{0.5}{\milli\watt}$). \textit{A} and \textit{B} are
exciton-like PR features centred at
$E_{\mathrm{A}}\approx\SI{1.19}{\electronvolt}$ and
$E_{\mathrm{B}}\approx\SI{1.63}{\electronvolt}$, respectively. The solid red
curve represents a numerical fit with the expression from Eq.~\ref{eq1}. 
Inset: a
sketch of the fundamental
exciton-like optical transitions at the K/K' points in bismuthene.}
\label{fig2}
\end{figure*}

\section*{Results I: Electronic band structure and single-particle band gap}

The bismuthene monolayers used in our optical experiments were grown by
molecular beam epitaxy using a Knudsen cell as Bi source and SiC(0001) as
supporting substrate (for details see Ref.~\onlinecite{Reis2017} and
\textbf{Methods}). The high quality of the resulting layers is best
visualized by scanning tunneling microscopy (STM). The topographic overview scan
in Fig.~\ref{fig1}{\bf a}, obtained by constant-current STM, shows an excellent Bi
coverage of $\sim87\%$. The inset depicts a detailed view 
on the honeycomb arrangement of the Bi atoms, ultimately confirming successful
film growth.

For a detailed discussion of the optical excitations and related exciton
physics it is mandatory to first develop a good understanding of the underlying
single-particle electronic properties of Bi/SiC. Detailed mapping of the valence
bands has already been performed by angle-resolved photoelectron spectroscopy
(ARPES) and found to be in convincing agreement with 
density-functional theory (DFT) \cite{Reis2017}. However, here we are interested
in both occupied and unoccupied bands and their relative energy separation which is
relevant for the optical transitions. This can be probed by STS which
records the differential tunneling conductivity as a measure of the electronic local density
of states (LDOS).

Fig.~\ref{fig1}{\bf b} displays the STS spectrum measured at the center of the
bismuthene sample where it shows its highest structural quality. The most
prominent feature of the spectrum is the vanishing LDOS between the energy
positions marked by $V_1$ and $C_1$. This region is identified as the
fundamental band gap, with the spectral onsets $V_1$ and $C_1$ representing the
valence band maximum (VBM) and conduction band minimum (CBM), respectively. For
a quantitative analysis the onset energy $V_1$ is evaluated by extrapolating the
valence band edge to the zero conductance plateau (see Supplementary
Information), an established method to determine gap edges in STS. The same
procedure is applied to the CBM at $C_1$. From the energy separation of $C_1$ and
$V_1$ we infer an experimental single-particle band-gap $E_{\mathrm{ind}}$ of
$\SI{0.80}{\electronvolt} \leq E_{\mathrm{ind}} \leq \SI{0.96}{\electronvolt}$.

In theory, the established method for a reliable estimate of the single-particle
gap is the $GW$ approximation, which expresses the electronic self-energy
$\Sigma$ by the electron propagator $G$ and the screened Coulomb
interaction $W({\bf k},\omega)$ via $\Sigma =
i{GW}$~\cite{Hedin1965,Strinati1982,Strinati1988}. This goes beyond DFT-type
approaches which are designed to yield accurate total (ground-state) energies but are not
devised for extracting single-particle energies~\cite{Kohn1965} and often tend to 
underestimate band gaps (see Fig.~\ref{fig1}{\bf c}, red curves for an example).

The $GW$ band structure for Bi/SiC (see \textbf{Methods} for the details) is
shown in Fig.~\ref{fig1}{\bf c} as solid blue lines. It is characterized by strongly
Rashba-split bands at the K (K') valleys and an indirect electronic band gap
between the VBM at K and the CBM at $\Gamma$ of $E_{\mathrm{ind}} = \SI{0.97}{\electronvolt}$, in
excellent agreement with our experimental gap value. The $GW$ band structure
accounts also for some other features in our experimental STS data, as indicated
by the dashed orange lines
connecting Figs.~\ref{fig1}{\bf b} and ~\ref{fig1}{\bf c}. For instance, the kink in
the measured LDOS at approximately $\SI{-0.5}{\electronvolt}$ (marked by $V_2$)
matches almost perfectly with the maximum of the lower Rashba-split $GW$ valence
band. Likewise, the experimental kink feature $C_2$ agrees well with the
K-point minimum of the $GW$ conduction band. Finally, and as
of immediate relevance for the optical transitions, the direct single-particle
gap at the K (K') valley amounts to $E_{\mathrm{direct}} =
\SI{1.36}{\electronvolt}$ in our $GW$ calculation.

\section*{Results II: Optical spectroscopy, optical gap and excitonic resonances}

Next, we expand our study to the optical response, i.e., to the
two-particle excitations. Experimentally, bismuthene's fundamental optical
excitation spectrum is probed via highly spatially-resolved photo-modulated
reflectivity (PR). PR is a very sensitive probe that has been successfully applied to study the excitonic spectrum of atomically thin crystals \cite{Zelewski17, Kopaczek22}, since it directly probes the derivative of the absorption spectrum with respect to energy, in stark contrast to direct reflectivity techniques.

The experimental setup is sketched in Fig.~\ref{fig2}{\bf a}
(see also \textbf{Methods}). All measurements were performed at room-temperature
and with the sample kept under an inert gas atmosphere to prevent surface
oxidation of the Bi monolayers. ARPES measurements performed before and after
exposure to inert gas confirm that this sample storage method keeps bismuthene
chemically intact over weeks (see also Supplementary Information), also owing to 
the van der Waals-nature of the bismuthene surface. The resulting
optical excitation spectrum is displayed in Fig.~\ref{fig2}{\bf b} (see also Supplementary Information). It is dominated by two absorption-like PR spectral features \textit{A} and \textit{B},
located at $\SI{\sim1.19}{\electronvolt}$ and $\SI{\sim1.63}{\electronvolt}$,
respectively. In analogy to other two-dimensional semiconducting
materials~\cite{Mak2010} and, in particular, taking into account the
aforementioned agreement between theory and experiment concerning the
single-particle properties, it is most reasonable to assign the captured optical
transitions to Coulomb-interacting electron-hole excitations in the vicinity of
K and K' points of the Brillouin zone (see inset in Fig.~\ref{fig2}{\bf b}).


\begin{figure*}[!t]
\centering
\includegraphics[width=\textwidth,angle=0,clip=true]{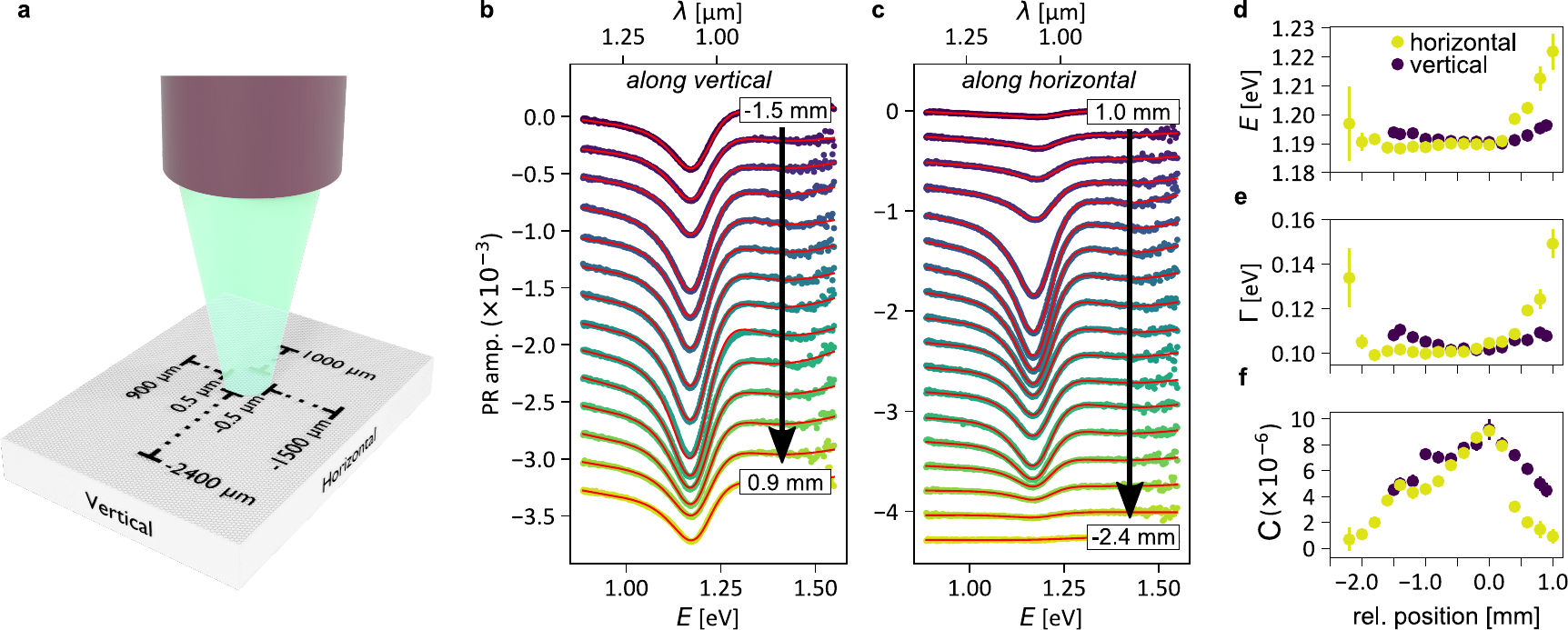}
\caption{{\bf Spatial modulation of the optical response of Bi/SiC.} 
{\bf a} 
Sketch of the scanning movement along a vertical and horizontal line on Bi/SiC. 
{\bf b,c} 
Evolution of the PR resonance of the \textit{A}-peak along the vertical and
horizontal scanning line, respectively. Solid red lines represent the fitting
curves (Eq.~\ref{eq1}).
{\bf d,e,f} 
The energy of the \textit{A} exciton transition, broadening, and
the PR amplitude, respectively. (Pulsed laser source,
$E_\mathrm{laser}\sim\SI{1.65}{\electronvolt}$, average pump power:$\SI{\sim0.34}{\milli\watt}$ , pump fluence:
$\SI{\sim0.57}{\milli\joule\per\centi\meter\tothe{2}}$). Error bars are the numerical fit uncertainties. The PR amplitude in horizontal direction correlates well with the bismuthene sample quality (see Supplementary Information).}
\label{fig3}
\end{figure*}

For a more quantitative data analysis we employ an expression for the PR
response of excitons based on an adapted model suggested by
Shanabrook~\textit{et al.}~\cite{Shanabrook1987} (see Supplementary Information). Within
this approach, the PR feature acquires the form of the first derivative of the
dielectric function:
\begin{equation}
	\frac{\Delta R}{R} = \mathfrak{R}\left(C_{\mathrm{j}} e^{i\theta}(E-E_{\mathrm{j}}+i\Gamma_{\mathrm{j}})^{-2}\right) \, .
	\label{eq1}
\end{equation}
Here, $C_{\mathrm{j}}$ and $\theta$ are the resonance amplitude and phase,
$E_{\mathrm{j}}$ and $\Gamma_{\mathrm{j}}$ are the energy and broadening
parameters of the optical transition, respectively. The fitting procedure to the
PR spectrum in Fig.~\ref{fig2}{\bf b} yields $E_{\mathrm{A}} = (1.191 \pm
0.004)~\si{\electronvolt}$ and $E_{\mathrm{B}} = (1.625 \pm
0.004)~\si{\electronvolt}$, $\Gamma_{\mathrm{A}} = (100 \pm
4)~\si{\milli\electronvolt}$ and $\Gamma_{\mathrm{B}} = (60 \pm
4)~\si{\milli\electronvolt}$. It is worth noting that both transitions show
substantial spectral broadening, which we attribute to effects
arising from, {\it e.g.}, sample inhomogeneity within the probed
area and interaction between excitons and phonons in these room-temperature
measurements~\cite{Marini2008,LouieTMD2013,PhysRevLett.115.119901}.

The high spatial resolution of the PR experiment allows us to probe our sample
from the coherent crystal regime in the center of the Bi monolayer to a more
disordered regime at its edges. As schematically depicted in Fig.~\ref{fig3}{\bf
a}, we follow the signature of the \textit{A}-resonance as the sample is scanned
along the indicated vertical and horizontal lines. The corresponding evolution
of the PR spectrum is presented in Figs.~\ref{fig3}{\bf b,c}. The fitting
procedure reveals that the transition energy (Fig.~\ref{fig3}{\bf d}),
broadening (Fig.~\ref{fig3}{\bf e}), and the PR amplitude (Fig.~\ref{fig3}{\bf
f}) indeed vary only weakly along the $\SI{2}{\milli\meter}$-long scanning line,
suggesting the fairly good quality of our bismuthene layer over a
macroscopically large spatial extension. The layer quality deteriorates close to
the bismuthene edges, which is clearly reflected by the diminishing PR amplitude
as well as the increased spectral broadening of the resonance. This observation
is in line with local structure characterization of our bismuthene films by STM
and low-energy electron diffraction (LEED) which reveal reduced crystalline order 
towards the sample edges (see Supplementary Information). The film
inhomogeneity from center to edge is most likely related to a temperature
gradient during epitaxial growth. Independent of the detailed mechanism, the marked
correlation between PR signal and local crystalline quality is a clear
indication that the observed PR resonances are indeed intrinsic to the pristine bismuthene layer.

Turning now to the theoretical picture of the optical response, first we wish to point
out that the many-body effects embedded in the single-particle band gap versus
those contained in the optical excitation spectrum are of different nature. Even
though electronic interactions are already relevant for the self-energy of the
single-particle excitations, as, {\it e.g.}, captured by our $GW$ approximation above,
the optical spectrum is additionally affected by the Coulomb coupling
between electrons and holes generated by the photo-excitation, lying at the very
heart of exciton physics. For a theoretical description of the two-particle properties
and the excitation spectrum including electron-hole interactions we solve the
Bethe-Salpeter equation (BSE) for the electron-hole 
amplitude~\cite{Hanke1975,Hanke1979,Rohlfing1998}:
\begin{equation}
(E_{c{\bf k}} - E_{v{\bf k}})A_{vc{\bf k}} + \Sigma_{v'c'{\bf k}'}\langle vc{\bf
k} |K_{eh} | v'c'{\bf k}'\rangle A_{v'c'{\bf k}'} = \Omega A_{vc{\bf k}} \,,
	\label{eq2}
\end{equation}
where the electronic excitations are given in the basis of
electron-hole pairs with quasi-particle energies $E_{v{\bf k}}$ and $E_{c{\bf
k}}$ in the valence and conduction bands. The $A_{vc{\bf k}}$ are the
coefficients of the excitons in the electron-hole basis and $\Omega$ are the
eigen-energies. The kernel $K_{eh}$ accounts for the screened Coulomb
interaction between electrons and holes, and the exchange interaction, including
also local field effects which are due to the optical excitations of the
periodically arranged atomic orbitals in Bi/SiC~\cite{Hanke1979b}.


\begin{figure*}
\centering
\includegraphics[width=\textwidth,angle=0,clip=true]{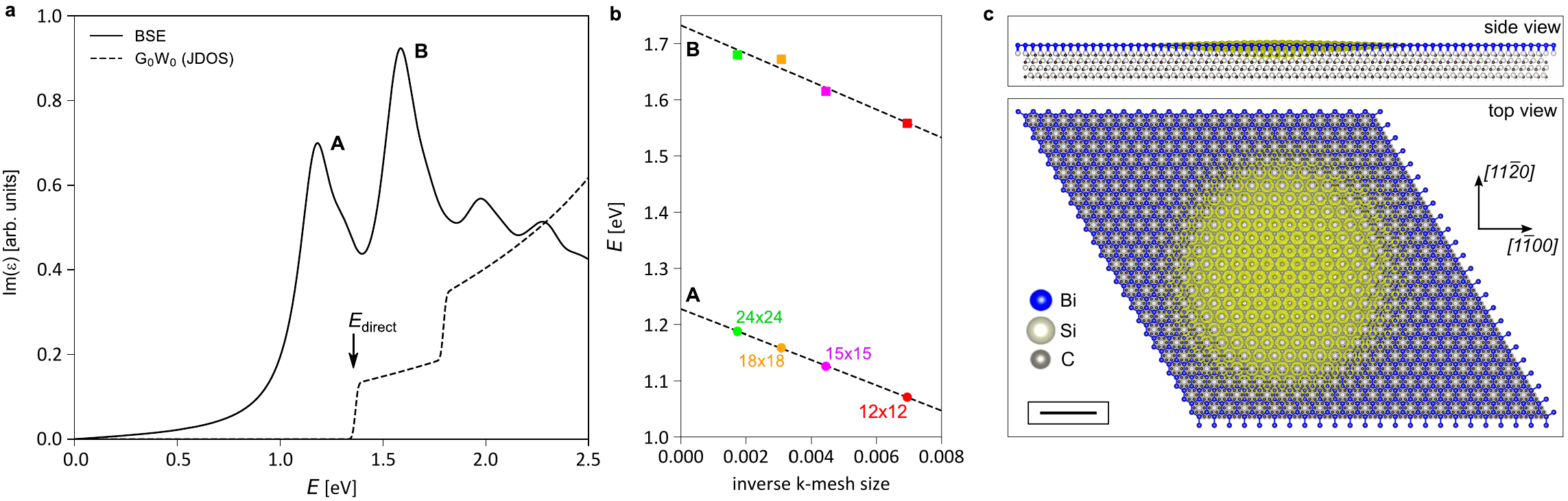}
\caption{{\bf Ab-initio $GW$+BSE results.} 
{\bf a} 
The solid line displays the absorption spectrum of Bi/SiC inferred from the imaginary part of the dielectric
function from the BSE calculation.
The dashed line displays the joint density of states from the $GW$ calculation, where the
electron-hole interaction is neglected.
%
{\bf b} 
Convergence of the \textit{A}~and \textit{B}~exciton peaks, as a function of the
inverse of the number of $k$-points. The intersection of the gray dashed line
with the $y$-axis gives the result ideally obtained from an infinitely dense
$k$-mesh.
{\bf c} 
Lateral and top views of the \textit{A}~exciton wave-function. The radius of
the exciton is $\sim\SI{3}{\nano\meter}$.} \label{fig4}
\end{figure*}

The absorption spectra calculated with and without electron-hole interactions
are displayed in Fig.~\ref{fig4}{\bf a}. From these calculations, we can extract
an excitonic binding energy of $\SI{\sim0.15}{\electronvolt}$ for the lowest
excitonic peak \textit{A}, estimated as the energy difference
between the independent particle $GW$ absorption gap $E_{\mathrm{direct}}$ indicated by the arrow and the energy position of peak \textit{A}. The second
evident peak \textit{B} is shifted by an energy comparable to the Rashba
splitting of the valence states at the K and K' valleys. This results in a
significant separation between the \textit{A} and \textit{B} excitonic peaks,
and the eventual merging of the latter into the continuum of electron-hole
excitations (forming a Fano resonance depending on coupling strength between
exciton \textit{B} and continuum).

We find that the position of the theoretical \textit{A} and \textit{B} peaks is highly sensitive to the number of $k$-points employed in the BSE calculation.
Fig.~\ref{fig4}{\bf b} shows the convergence for several $k$-meshes,
highlighting a linear scaling towards
$E_{\mathrm{A}}\approx\SI{1.23}{\electronvolt}$ for an infinitely dense
sampling, in excellent agreement with its measured value ($\SI{1.19}{\electronvolt}$).
For exciton peak \textit{B}, the extrapolation yields a theoretical value of 
$E_{\mathrm{B}}\approx\SI{1.73}{\electronvolt}$, still satisfyingly close to 
the experimental result ($\SI{1.63}{\electronvolt}$).

The aforementioned convergence issue arises from the extended nature of the
Wannier-Mott type excitons in Bi/SiC. In Fig.~\ref{fig4}{\bf c}, we give side and
top views of the \textit{A} peak excitonic wave-function obtained from our
calculations. The exciton radius of $\SI{\sim 3}{\nano\meter}$ extends over
several lattice constants, resulting from the strong localization in reciprocal
space around the K and K' valleys. This, in turn, requires a dense $k$-mesh to
reduce the spurious Coulomb repulsion between periodic excitons. Remarkably,
Fig.~\ref{fig4}{\bf c} shows also how the exciton wave-function is not strictly
localized inside the Bi monolayer, but rather extends for at least two atomic
layers into the SiC substrate. This further evidences the pivotal role
played by the substrate in bismuthene at the single- and two-particle levels:
SiC is not only responsible for the orbital filtering mechanism and the
huge topological gap~\cite{Reis2017,GangLi2018}, but also provides an important
screening channel for the Coulomb interaction.

\section*{Discussion}

In conclusion, we report the observation of excitons in a large-gap,
atomically-thin QSH insulator at room temperature. We capture two prominent
transitions in the many-particle response of bismuthene which we quantitatively
associate with excitonic transitions from the Rashba-split valence bands. It is
important to emphasize that the many-particle response of bismuthene differs
fundamentally from what is known from the vastly studied 2H-TMDCs. First,
whereas in the latter materials the low-energy interband transitions are captured
by two decoupled gapped chiral Dirac fermion models~\cite{Zhang2018,Cao2018}
with the same winding number in each valley (with opposite sign between
K/K'-valley), in bismuthene, low-energy interband transitions are described by
chiral models with different winding numbers in each valley. Therefore, in
contrast to the strict valley-locked optical selection rules that govern
2H-TMDCs, bismuthene is sensitive to controlling the coupling to $\sigma^+$- and
$\sigma^-$-polarized light by reordering electronic bands with external electric
and magnetic fields~\cite{Xu2020} (see also Supplementary Information). The latter provides a clear
technological advantage over TMDCs for engineering electro-optical devices based
on atomically thin materials. Second, in bismuthene the excitonic \emph{and} the
topological physics originate from the very same electronic states, {\it i.e.},
they establish a direct link between excitonic physics and topology, which has
only been proposed theoretically so far (see for example,
Ref.~\cite{FabrizioPRB}). We foresee that our findings trigger a plethora of
future experimental investigations, for example related to the helical excitonic
transport in topological edge modes~\cite{MacDonald2017}, finding and assessing
``global" topological properties of the QSH phase via optical band-to-band
transitions and selection rules (see also Supplementary Information), topological
polaritonics~\cite{janot2016topological}, and other exotic many-body quantum
phases.
\\


\noindent{\bf Methods}\\
\begin{scriptsize}
\noindent{\bf Sample preparation --} 
The bismuthene sample was grown by
molecular beam epitaxy using a Knudsen cell as Bi source and SiC(0001) as
supporting substrate. During growth, the temperature of the SiC substrates was
controlled via direct current resistive heating. The characteristic of this
heating method is a~temperature gradient of $\SI{\sim20}{\degreeCelsius}$ along
the longitudinal extent of the sample (the horizontal direction). As bismuthene
growth is sensitive to such a narrow temperature window, we observe a~strong
correlation of bismuthene film coverage with the spatial position in the
horizontal direction. Based on low energy electron diffraction (LEED), the
bismuthene film quality can be macroscopically checked. High intensity of the
$\sqrt{3}\times\sqrt{3}$ spots associated with the bismuthene reconstruction on
SiC(0001) and low diffuse background intensity provide a qualitative measure of
film quality (see Supplementary Information).

\noindent{\bf Optical setup --} 
In the PR experiment (Fig.~\ref{fig2}{\bf a} and Fig.~\ref{fig3}{\bf a,c}), the
sample is kept under a N$_2$ inert gas atmosphere ($<0.1$~ppm H$_2$O;
$<0.1$~ppm O$_2$) and mounted on an x-y-z mechanical stage, allowing for selecting
the
probed sample's area with a $\SI{\sim100}{\nano\meter}$ accuracy in each spatial
direction. The infinity-corrected microscope objective ($0.42$ numerical
aperture and $\SI{20}{\milli\meter}$ working distance) used in the PR experiment
operates in the spectral range $\SIrange{0.4}{1.8}{\micro\meter}$ with
$\times20$~magnification. The focal plane of the modulation laser beam is inspected
through the imaging setup with a CCD camera showing the sample surface, the
white light spot, and the laser spot. Since the modulation laser spot is roughly
$\SI{1}{\micro\meter}$ in diameter, it defines the total area of a sample tested
in a single experimental run. White light comes from the Halogen lamp. A
mechanical chopper modulates the laser beam with a $50\%$:$50\%$ duty cycle at
an $f_{\mathrm{m}}$ frequency depending on the utilized photon detector: (i)~the
$f_{\mathrm{m}}=\SI{10}{\hertz}$ for the Si diode, sensitive in the
$\SIrange{0.3}{1.0}{\micro\meter}$ spectral range, or
(ii)~$f_{\mathrm{m}}=\SI{270}{\hertz}$ for the InGaAs photo-diode, covering
the spectral range of $\SIrange{0.9}{1.5}{\micro\meter}$. Several types of laser
sources are employed: (a)~the continuous-wave (CW) neodymium-doped yttrium
aluminum garnet laser emitting at $\SI{532}{\nano\meter}$ photon wavelength;
(b)~the ultrafast Ti: Sapphire oscillator, providing a train of
$\sim\SI{140}{\femto\second}$-long-pulses at 76 MHz repetition frequency,
spectrally tunable in the range of $\SIrange{0.7}{0.95}{\micro\meter}$ or
$\SIrange{0.35}{0.475}{\micro\meter}$ (second harmonic generation), and (c)~the
synchronously-pumped optical parametric oscillator providing a train of
$\sim\SI{200}{\femto\second}$-long-pulses spectrally tunable in the range of
$\SIrange{0.5}{0.76}{\micro\meter}$ or $\SIrange{0.97}{1.1}{\micro\meter}$. A
spectral light analysis is provided by a spectrometer consisting of
a~$\SI{0.3}{\meter}$-focal-length grating monochromator ($\SI{0.2}{\nano\meter}$
ultimate spectral resolution) and two detectors mentioned above. The spectral
resolution is limited to an arbitrarily chosen monochromator scanning step in
the PR experiment, which is $\SI{\sim4}{\milli\electronvolt}$. The lock-in
amplifier reads the detector signal at the $f_{\mathrm{m}}$ carrier frequency.

\noindent{\bf Theoretical framework --}
The approach to determine and quantify the importance of the electron-hole
interaction, and corresponding excitonic effects, in Bi/SiC employed the
following methods (in this hierarchical order): $DFT \rightarrow GW \rightarrow BSE$. Density Functional Theory (DFT) is used to understand the ground-state
properties, with the interacting many-electrons problem being mapped onto a
non-interacting electrons problem. Spectroscopic properties, however, involve
excited particles above the ground state and a correct step in this direction
can be done by considering the many-body electron-electron interaction in the
framework of $GW$ approximation. Starting
from DFT single-particle energies, one can compute the Quasi-Particle (QP)
energies by taking into account the self-energy $\Sigma$ expressed through the
single-particle Green's function $G$ and the screened Coulomb interaction $W$ (here computed within the random phase approximation RPA including local field effects).
This approximation is known to be correct at about 0.1 eV~\cite{Ugeda2014}, which makes our $GW$ results overall consistent with all our STS spectra shown in the manuscript.
Finally, on top of the $GW$ results, the effect of the electron-hole
interactions are included using the Bethe-Salpeter Equation (BSE) and the
excitonic two-body wave-function is written via a product of occupied and
unoccupied orbitals. The results here presented have been obtained using the
simulation packages Quantum Espresso~\cite{Giannozzi2009} and
YAMBO~\cite{Sangalli2019,MARINI20091392}. Norm-conserving pseudo-potentials are used to quantify the electron-ion interactions and the Perdew-Burke-Ernzerhof (PBE)
functional~\cite{Perdew1996} is used for the exchange-correlation potential. Wave-functions are expanded in plane-waves with an energy cut-off of 40 Ry, large enough to guarantee converged results. The Brillouin zone is then sampled with a 12$\times$12$\times$1 grid for DFT ionic relaxation and non
self-consistent calculations, but it has been increased up to
24$\times$24$\times$1 $k$-points for many-body calculations. Spin-orbit coupling is included self-consistently. QP energies are then computed using a single shot $G_0 W_0$ approximation.




\end{scriptsize}

\bibliographystyle{naturemag}


\ \\

\noindent{\bf Acknowledgements}\\
\begin{scriptsize}
The authors acknowledge insightful discussions with B. Rosenow, B. Scharf and C. Franchini. The research leading to these results has received funding from the European Union's Horizon 2020 research and innovation program under the Marie Sk{\l}odowska-Curie Grant Agreement No. 897276. We gratefully acknowledge the Gauss Centre for Supercomputing e.V. (https://www.gauss-centre.eu) for funding this project by providing computing time on the GCS Supercomputer SuperMUC-NG at Leibniz Supercomputing Centre (https://www.lrz.de). We are grateful for funding support from the Deutsche Forschungsgemeinschaft (DFG, German Research Foundation) under Germany's Excellence Strategy through the W\"urzburg-Dresden Cluster of Excellence on Complexity and Topology in Quantum Matter ct.qmat (EXC 2147, Project ID 390858490) as well as through the Collaborative Research Center SFB 1170 ToCoTronics (Project ID 258499086).
P.~H. acknowledges support from Polish National Science Center within the Etiuda 8 scholarship (Grant No. 2020/36/T/ST5/00511). M.~S. acknowledges support from the Polish National Science Center within the Opus 18 Grant No.2019/35/B/ST5/04308 and the Polish National Agency for Academic Exchange (PPI/APM/2018/1/00031/U/001).
The Flatiron Institute is a division of the Simons Foundation.\\

\end{scriptsize}

\noindent{\bf Author contributions}\\
\begin{scriptsize}
C.S. and R.C. conceived the project. M.S., P.H, P.W. and {\L}.D. performed the photo-modulated reflectivity experiments and analyzed the resulting data with help from S.H. and C.S.. R.S. has grown the sample and conducted the STM/STS and ARPES characterization. R.S. and F.R. conducted the XPS characterization. A.C. and D.D.S. performed the ab-initio $GW$+BSE calculations, with input from R.T. and W.H.. C.S., R.C., W.H. and D.D.S. wrote the manuscript with input from all coauthors.\\

\end{scriptsize}

\noindent{\bf Additional information}\\
\begin{scriptsize}
{\bf Extended data} is available for this paper at https://xxxx\\
{\bf Supplementary information} The online version contains supplementary material available at https://xxx\\
{\bf Correspondence and requests for materials} should be addressed to Christian Schneider, Ralph Claessen and Domenico Di Sante.\\
\end{scriptsize}

\noindent{\bf Competing financial interests}\\
\begin{scriptsize}
The authors declare no competing financial interests.
\end{scriptsize}

\end{document}